# EC-FORC: A New Cyclic Voltammetry Based Method for Examining Phase Transitions and Predicting Equilibrium

I. Abou Hamad[a], D. T. Robb[b], M.A. Novotny[a,c], P.A. Rikvold[d,e,f]

[a] HPC$^2$, Center for Computational Sciences, Mississippi State University,
Mississippi State, MS 39762-5167, USA
[b] Department of Physics, Box 5721, Clarkson University, Potsdam, NY 13699, USA
[c] Department of Physics & Astronomy, Mississippi State University,
Mississippi State, MS 39762-5167, USA
[d] Center for Materials Research and Technology and Department of Physics,
Florida State University, Tallahassee, FL 32306-4350, USA
[e] School of Computational Science, Florida State University,
Tallahassee, FL 32306-4120, USA
[f] National High Magnetic Field Laboratory, Tallahassee, FL 32310, USA

We propose a new, cyclic-voltammetry based experimental technique that can not only differentiate between discontinuous and continuous phase transitions in an adsorbate layer, but also quite accurately recover equilibrium behavior from dynamic analysis of systems with a continuous phase transition. The Electrochemical first-order reversal curve (EC-FORC) diagram for a discontinuous phase transition (nucleation and growth), such as occurs in underpotential deposition, is characterized by a negative region, while such a region does not exist for a continuous phase transition, such as occurs in the electrosorption of Br on Ag(100). Moreover, for systems with a continuous phase transition, the minima of the individual EC-FORCs trace the equilibrium curve, even at very high scan rates. Since obtaining experimental data for the EC-FORC method would require only a simple reprogramming of the potentiostat used in conventional cyclic-voltammetry experiments, we believe that this method has significant potential for easy, rapid, in-situ analysis of systems undergoing electrochemical deposition.

## Introduction

Studies of atomic-scale dynamics in electrochemical deposition have been made possible by recent technological developments (1). These new experimental methods have increased the need for new methods for computational analysis of experimental adsorption dynamics. One such analysis method is the electrochemical first-order reversal curve (EC-FORC) method (2). In this paper, we briefly describe the EC-FORC method and provide additional confirmation of its validity by applying it to a mean-field model of electrochemical adsorption with a discontinuous phase transition.

The EC-FORC method for a lattice-gas model of electrochemical adsorption consists of saturating the adsorbate coverage in a strong positive electrochemical potential and, in each case starting from saturation, decreasing the potential to a series of progressively more negative 'reversal potentials', and subsequently increasing the potential gradually back to the saturation value. The method can also be employed, starting from zero coverage (desaturation) and using a range of progressively more positive reversal potentials. In this latter version, the EC-FORC method is a generalization of the "switching-potential technique" for detecting discontinuous phase transitions progressing via nucleation and growth through a current extremum along the potential-reversal path, which was pioneered by Fletcher and collaborators (3). The EC-FORC method can be implemented in a cyclic-voltammetry experiment by controlling the electrochemical potential through the electrode potential and cycling between the saturation (or desaturation) electrode potential and progressively more negative (or more positive) reversal electrode potentials, $E_r$. This should require only a simple reprogramming of the potentiostat. In the simulations shown here, we start at saturation and use progressively more negative reversal potentials. After calculating the coverage from the value of the current at each point (2), and the corresponding value of the electrochemical potential from the value of the electrode potential, a family of EC-FORCs $\theta(\bar{\mu}_r, \bar{\mu}_i)$ is produced, where $\theta$ is the coverage, $\bar{\mu}_r$ is the electrochemical potential corresponding to $E_r$, and $\bar{\mu}_i$ is the instantaneous electrochemical potential during the increase back to saturation. The EC-FORC distribution, which is defined as

$$\rho = -\frac{1}{2}\frac{\partial^2 \theta}{\partial \bar{\mu}_r \partial \bar{\mu}_i} = \frac{1}{2\gamma e (d\bar{\mu}_i / dt)} \frac{\partial I(\bar{\mu}_r, \bar{\mu}_i)}{\partial \bar{\mu}_r} \quad , \qquad [1]$$

where $I$ is the current density, $\gamma$ is the electrosorption valency, and $e$ is the electronic charge unit, can then be calculated by fitting a polynomial surface to a grid of data points $\theta(\bar{\mu}_r, \bar{\mu}_i)$.

## Monte Carlo Simulations of a Lattice-gas Model

In this section we briefly describe the model and some results that were previously published in Ref. (2). Kinetic Monte Carlo (KMC) simulations of a lattice-gas model, where a Monte Carlo (MC) step corresponds to an attempt to cross a local free-energy barrier, were used to simulate the kinetics of electrochemical systems with first- (4,5) or second-order (6,7) phase transitions in two dimensions. A grand-canonical effective lattice-gas Hamiltonian (6-14)

$$H = -\sum_{i<j} \phi_{ij} c_i c_j - \bar{\mu} \sum_{i=1}^{N} c_i \qquad [2]$$

is used to describe the energy associated with a specific adsorbate configuration. Here, $\sum_{i<j}$ is a sum over all pairs of sites, $\phi_{ij}$ are the lateral interaction energies between particles on the $i$th and $j$th lattice sites, $\bar{\mu}$ is the electrochemical potential, and $N=L^2$ is the total number of lattice sites for an $L$x$L$ square lattice with periodic boundary conditions. The local occupation variable $c_i$ is 1 if site $i$ is occupied and 0 otherwise. In addition to adsorption/desorption steps, we include diffusion steps that have a free-energy barrier comparable to the adsorption/desorption free-energy barrier (7).

Independent of the diffusion, attractive nearest-neighbor interactions ($\phi_{ij}>0$) produce a discontinuous phase transition equivalent to that which occurs in systems undergoing underpotential deposition (UPD) (4,5,15). In contrast, repulsive long-range interactions ($\phi_{ij}<0$) and nearest-neighbor exclusion produce a continuous phase transition, equivalent to that in Br or Cl adsorption on a Ag(100) surface (6,7,9,16,17).

Continuous Phase Transitions

The EC-FORCs and the EC-FORC diagram for a system with a continuous phase transition are shown in figures 1a and 1c, respectively. The corresponding current densities are shown in figure 1b. Figure 1a also shows the EC-FORC minima and a plot of the equilibrium behavior of the system. Notice that the minima lie directly on the equilibrium isotherm. This is because such a system has only one stable state for any given value of the electrochemical potential, defined by the continuous equilibrium curve. The minima follow the equilibrium behavior closely, even at much higher scan rates of the electrochemical potential (Fig. 2). The positive values of $\rho$ in the EC-FORC diagram (Fig. 1c) result from the relaxation toward the equilibrium isotherm, at a rate which increases with the distance from equilibrium. The unusual structure in the EC-FORC diagram in Fig. 1c, which resembles a 'Florida-shaped peninsula' centered around the EC-FORC that passes closest to the critical coverage point (shown in bold in Figs. 1a and 1c), suggested to us that we should examine the dynamic approach to equilibrium in the FORCs above and below this FORC. As seen in the inset to Fig. 2, at a higher scan rate (given in milli electron-volt / Monte Carlo step per site) the difference in this approach to equilibrium is indeed clearly visible, and can be related to the phenomenon of jamming, as discussed in Ref. (2). This sensitivity of the EC-FORC method to the dynamics of phase transformation should make it a useful tool for investigation.

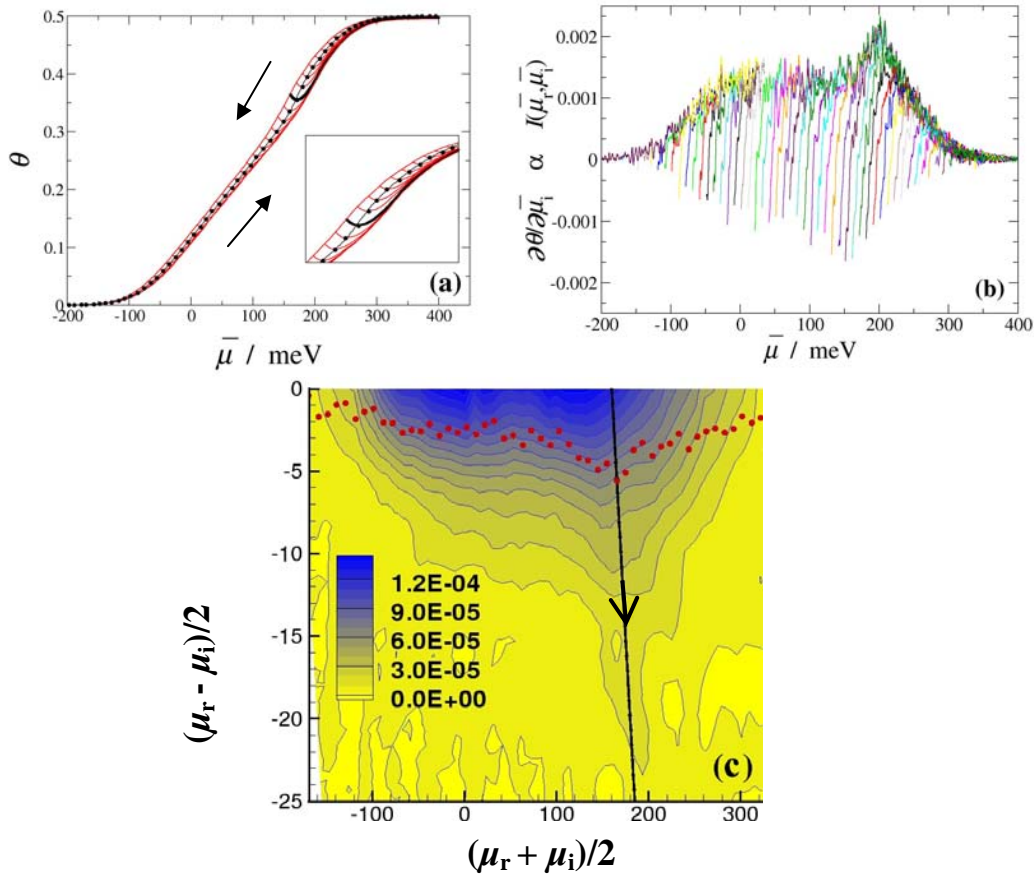

Figure 1. (Color online.) **(a)** EC-FORCs for a system with a continuous phase transition and a low scan rate (0.0003 meV/MCSS). The black dots are the minima of the individual EC-FORCs, and the line they trace is the equilibrium curve calculated independently. The bold black line is the EC-FORC that passes closest to the critical coverage. The inset shows an expanded view of the upper right-hand portion of the curve. **(b)** The voltammetric currents corresponding to the EC-FORCs in **(a)**. **(c)** EC-FORC diagram for the same system. No negative regions are seen, and a "peninsula" is centered around the bold black line equivalent to the bold curve in **(a)**. The filled circles are the EC-FORC minima.

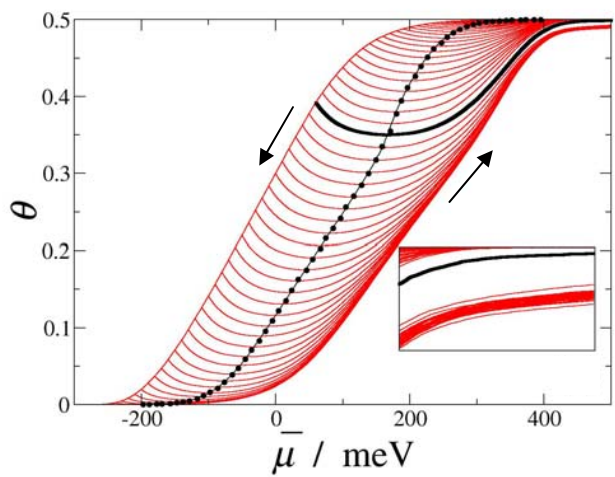

Figure 2. (Color online.) EC-FORCS for a system with a continuous phase transition and a high scan rate (0.01 meV/MCSS). The black dots are the minima of the individual EC-FORCS, and the line they trace is the equilibrium curve calculated independently. The inset shows an expanded view of the upper right-hand portion of the curve where the difference in approach to equilibrium due to the phenomenon of jamming is clearly visible.

Discontinuous Phase Transitions

Using the above Hamiltonian (Eq. 2) and attractive nearest-neighbor interactions in a lattice-gas model results in a system with a discontinuous phase transition similar to the one encountered in UPD. The EC-FORCs at a scan rate of 0.3 meV/MCSS are shown in figure 3a, with a vertical line indicating the position of the coexistence value of the electrochemical potential, $\bar{\mu}_c$, and circles showing the positions of the minima. Contrary to a simple Avrami's-law analysis for an infinitely large system, which shows the minima to lie at $\bar{\mu}_i = \bar{\mu}_c$ (18) (corresponding to the clean "current crossings" observed in Ref. (3)), the minima here show a 'back-bending' behavior. This behavior is due to a competition between a nearby metastable state and a distant stable state. This local instability causes the divergence of the EC-FORCs and is reflected as a negative region centered around the line that passes closest to the unstable point ($\theta = 0.5$) in the EC-FORC diagram (Fig. 3c).

The voltammetric currents corresponding to the EC-FORCS in figure 3a are shown in figure 3b. The currents corresponding to those EC-FORCS whose minima lie on the back-bending part of the curve traced by the black points in figure 3a each show a minimum along the reversal path. This effect is in agreement with the observation of current extrema by Fletcher et al. (3). However, the current extrema in this lattice-gas model are not very pronounced, and the negative value of $\rho$ is a more striking sign of the discontinuous phase transition.

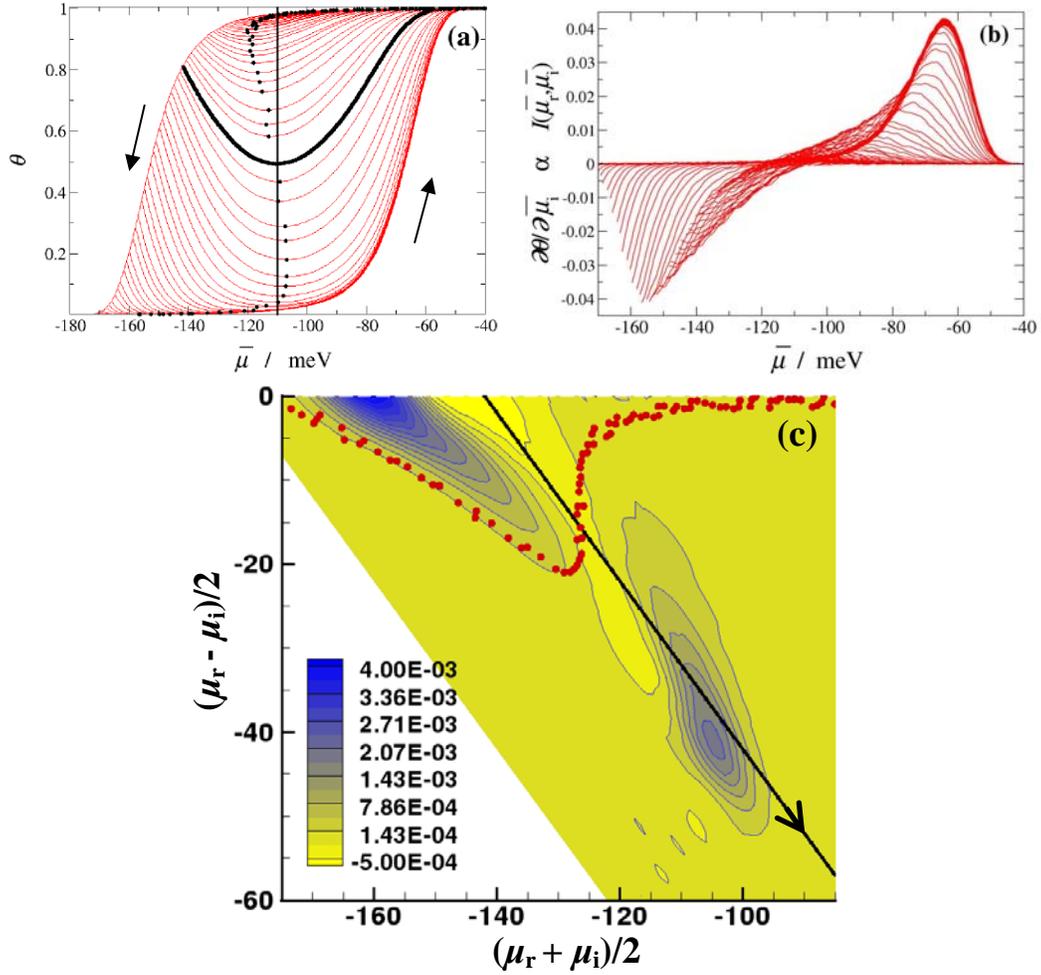

Figure 3. (Color online.) **(a)** EC-FORCs for a lattice-gas model of a system with a discontinuous phase transition. The vertical line shows the position of the coexistence electrochemical potential, and the filled circles are the minima of the EC-FORCs. The thick curve corresponds to the EC-FORC whose minimum lies nearest the coexistence electrochemical potential. **(b)** The voltammetric currents corresponding to the EC-FORCs in **(a)**. **(c)** EC-FORC diagram generated from the EC-FORCs in **(a)**. Note the negative region. The thick straight line corresponds to the EC-FORC marked as a thick curve in **(a)**, and the filled circles correspond to EC-FORC minima.

## Mean-field Model of a Discontinuous Phase Transition

We can describe the competition between the self-ordering and the potential explicitly using a time dependent mean-field model of the dynamics. The free energy is approximated by the quartic "Ginzburg-Landau" form,

$$F(\theta,\overline{\mu}(t)) = a\frac{(\theta-\theta_c)^4}{4} - b\frac{(\theta-\theta_c)^2}{2} - (\theta-\theta_c)(\overline{\mu}(t)-\overline{\mu}_c),\qquad[3]$$

where $\theta_c = 0.5$ is the coexistence coverage, and $\bar{\mu}_c = -110$ meV is the electrochemical potential at coexistence. In the noise-free (zero-temperature) case, the dynamics are given by

$$\frac{d\theta}{dt} = -\frac{1}{\gamma}\frac{dF}{d\theta} = -\frac{1}{\gamma}(a(\theta-\theta_c)^3 - b(\theta-\theta_c) - (\bar{\mu}(t) - \bar{\mu}_c)). \qquad [4]$$

The parameters for this mean-field model can be determined directly from an experimental or simulated family of EC-FORCs. On the line of minima of the EC-FORCs, $\frac{d\theta}{dt} = 0$, which implies $a(\theta-\theta_c)^3 - b(\theta-\theta_c) - (\bar{\mu}(t) - \bar{\mu}_c) = 0$. Differentiating this equation with respect to $\bar{\mu}$ and solving for $\frac{d\theta}{d\bar{\mu}}$ gives

$$\frac{d\theta}{d\bar{\mu}} = \frac{1}{3a(\theta-\theta_c)^2 - b} \quad \text{(along line of minima.)} \qquad [5]$$

For $\theta = \theta_c$, this yields $b = -\left(d\theta/d\bar{\mu}\big|_{\theta=\theta_c}\right)^{-1}$. The spinodal point along the line of minima occurs at the coverage $\theta_{sp}$, where $3a(\theta_{sp} - \theta_c)^2 - b = 0$, yielding $a = b/3(\theta_{sp} - \theta_c)^2$. The damping parameter $\gamma$ can be related to the 'coercive potential' $\bar{\mu}_{coer}$ (the reversal potential, $\bar{\mu}_r$, for which the coverage $\theta$ is closest to $\theta_c$), the slope $\frac{d\theta}{d\bar{\mu}}\bigg|_{\bar{\mu}=\bar{\mu}_{coer}}$, and the slope $\frac{d\bar{\mu}}{dt}$ as

$$\frac{d\theta}{d\bar{\mu}}\bigg|_{\bar{\mu}=\bar{\mu}_{coer}} = \frac{d\theta/dt\big|_{\theta=\theta_c}}{d\bar{\mu}/dt} = \frac{\bar{\mu}_{coer} - \bar{\mu}_c}{\gamma d\bar{\mu}/dt}, \qquad [6]$$

where we have used Eq. 4 in the last step. We determined in this manner the parameters from the family of FORCs in Fig. 3a as $b = 16.0$ meV, $\theta_{sp} = 0.92$, $a = 30.2$ meV $\bar{\mu}_{coer} = -153$ meV, $\frac{d\theta}{d\bar{\mu}}\bigg|_{\bar{\mu}=\bar{\mu}_{coer}} = -0.0385$ meV$^{-1}$, and $\gamma = 3720$ meV-MCSS. (Note that $d\bar{\mu}/dt$ must be taken as -0.3 meV/MCSS in applying Eq. 6 at the coercive potential.)

In figure 4a, we show the family of EC-FORCs obtained by numerical integration of the mean-field model. The line of minima follows the analytical curve defined by Eq. 5, as expected, and shows the 'back-bending', similar to the one seen for the lattice-gas model (Fig. 3a). Moreover, the existence of the negative region is reconfirmed in figure 4b, again similar to the one expected by the lattice-gas model (Fig. 3b). In addition, while simulations of the lattice-gas model with attractive interactions show a dependence of the line of minima on the sweep rate, in the mean-field model this line remains unchanged for different sweep rates. Preliminary simulations indicate that including thermal noise may further improve the agreement with the Monte Carlo results.

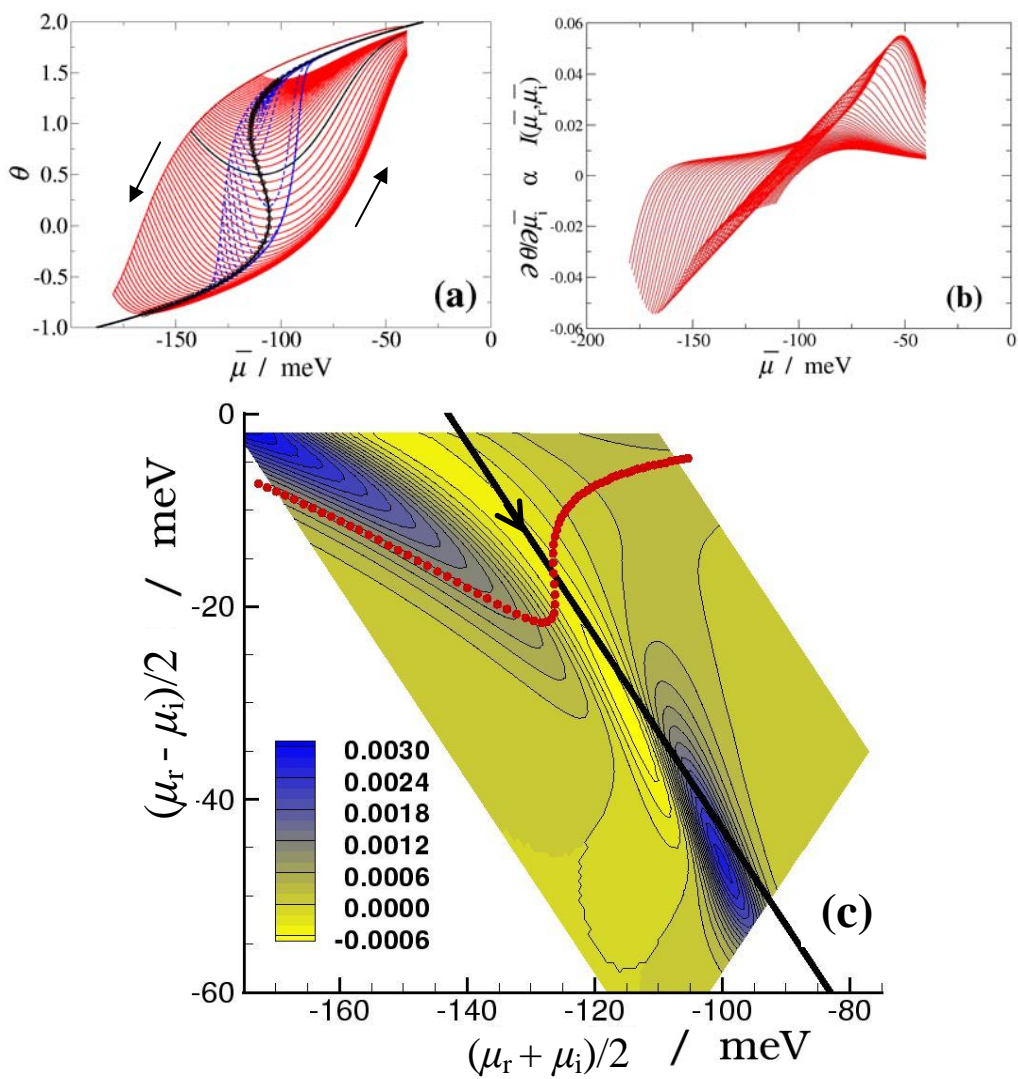

Figure 4. (Color online.) Results for the mean-field model. **(a)** Family of EC-FORCs. The solid lines show the EC-FORCs for the faster sweep rate (0.3 meV/MCSS), and the dashed lines show the EC-FORCs for a slower sweep rate (0.03 meV/MCSS). The thick, *S*-shaped curve is the analytical result (Eq. 5) for the line of minima, while the circles show the actual minima from the numerical integration. **(b)** The voltammetric currents corresponding to the EC-FORCs in (**a**). **(c)** EC-FORC diagram corresponding to the EC-FORCs in (a). The closed circles are the EC-FORC minima. Note the negative region. The overall appearance is very similar to Fig. 3c.

## Conclusions

We have previously shown the ability of the EC-FORC method not only to distinguish between phase transitions of different orders in electrochemical adsorption, but also to recover from dynamical data the equilibrium isotherm in systems undergoing a continuous phase transition. In this paper we have recapitulated the highlights of the method, and applied it to a time-dependent mean-field model of a continuous phase transition.. With simple reprogramming of the potentiostat, experimentalists can use the advantages of this method to study both the equilibrium and dynamic behaviors of electrochemical adsorption processes.

## Acknowledgments

The authors acknowledge useful correspondence with S. Fletcher. This work was supported by U.S. National Science Foundation Grant Nos. DMR-0240078 (Florida State University) and DMR-044405 (Florida State University and Mississippi State University) and DMR-0509104 (Clarkson University) and by ABSL Power Solutions, Inc. award No. W15P7T06CP408.